\documentclass[review,authoryear,12pt]{elsarticle}

 \usepackage{graphicx}

\usepackage{units}
\usepackage{amssymb}
 \usepackage{amsmath,amssymb,amsfonts,mathrsfs,amsthm}

 \usepackage{lineno}

\journal{Advances in Water Resources}

\begin{document}

\begin{frontmatter}


\title{Title\tnoteref{label1}}
\author{Caroline Dorn \fnref{label1}}

\title{Conditioning of stochastic 3-D fracture networks to hydrological and geophysical data}


\author[label1]{Niklas Linde}
\author[label2]{Tanguy Le Borgne}
\author[label2]{Olivier Bour}
\author[label2]{Jean-Raynald de Dreuzy}
\address[label1]{University of Lausanne, Faculty of Geoscience and the Environment, Applied and Environmental Geophysics Group, Switzerland}
\address[label2]{Universit\'e Rennes 1-CNRS, OSUR G\'eoscience Rennes, Campus Beaulieu, 35042 Rennes, France} 

\address{}
\begin{abstract}

The geometry and connectivity of fractures exert a strong influence on the flow and transport properties of fracture networks. We present a novel approach to stochastically generate  three-dimensional discrete networks of connected fractures that are conditioned to hydrological and geophysical data. A hierarchical rejection sampling algorithm is used to draw realizations from the posterior probability density function at different conditioning levels. The method is applied to a well-studied granitic formation using data acquired within two boreholes located 6~m apart. The prior models include 27~fractures with their geometry (position and orientation) bounded by information derived from single-hole ground-penetrating radar (GPR) data acquired during saline tracer tests and optical televiewer logs. Eleven cross-hole hydraulic connections between fractures in neighboring boreholes and the order in which the tracer arrives at different fractures are used for conditioning. Furthermore, the networks are conditioned to the observed relative hydraulic importance of the different hydraulic connections by numerically simulating the flow response. Among the conditioning data considered, constraints on the relative flow contributions were the most effective in determining the variability among the network realizations. 
Nevertheless, we find  that the posterior model space is strongly determined by the imposed prior bounds. Strong prior bounds were derived from GPR measurements and helped to make the approach computationally feasible. We analyze a set of 230 posterior realizations that reproduce all data given their uncertainties assuming the same uniform transmissivity in all fractures. The posterior models provide valuable statistics on length scales and density of connected fractures, as well as their connectivity. 
In an additional analysis, effective transmissivity estimates of the posterior realizations indicate a strong influence of the DFN structure, in that it induces large variations of equivalent transmissivities between realizations.
 The transmissivity estimates agree well with previous estimates at the site based on pumping, flowmeter and temperature data.

\end{abstract}


\end{frontmatter}


\section{Introduction}

The characterization of flow, storage, and transport properties in fractured rock formations is challenging \cite[e.g.,][]{Bonnet_2001,Berkowitz_2002,neuman2005trends}. The challenge resides mainly in the extreme heterogeneity of fractured formations, in which hydraulic conductivity varies over many orders of magnitudes within very short distances. Furthermore, flow and transport in fractured systems are highly organized, with flow channeling within fracture planes and preferential pathways within the connected fracture network \cite[e.g.,][]{smith1984analysis,neretnieks1990solute}. Direct measurements of hydraulic or geometrical properties of all individual fractures are impossible, and the data coverage in field investigations is typically very low compared with the underlying heterogeneity. One solution to this problem is to derive effective bulk hydraulic properties \cite[]{marechal2004use}, but this approach has been widely criticized, as it implies the existence of a homogenization scale that is smaller than the scale of investigation \cite[]{Berkowitz_2002,neuman2005trends}. Ongoing research focuses on the hydraulic and geometrical characterization of fracture networks (e.g., network permeability, fracture density, fracture length distribution), of individual fractures (e.g., position, location, orientation, transmissivity) and their connectivity \cite[e.g.,][]{Berkowitz_2002}.

Hydrological investigations of fractured rock aquifers often include time-consuming hydraulic, flowmeter and tracer tests \cite[e.g.,][]{Harvey_2000,Day-Lewis_2006,LeBorgne_2007}. These tests allow investigating hydraulic properties and dispersive phenomena in transmissive fracture networks that intersect the boreholes. Flowmeter and tracer tests provide effective parameter estimates for the investigated hydraulic connections, which might comprise many fractures with complex connectivity patterns \cite[e.g.,][]{Frampton}. Inferring hydraulic characteristics of the individual fractures is most often impossible, unless if conditions are favorable \cite[]{Novakowski}. Even extensive packer tests fail to resolve geometrical and hydraulic properties of individual fractures \cite[]{hao2007hydraulic}.

Previous geophysical field investigations in fractured rocks have demonstrated the value added of single-hole ground-penetrating radar (GPR) reflection data in providing constraints on the geometry of individual fractures \cite[e.g.,][]{Olsson_1992,Dorn_2012}. Properties, such as their location and orientation relative to the borehole, can be constrained from GPR images \cite[e.g.,][]{Olsson_1992,Grasmueck_1996,tsoflias2004vertical}. Compared with other geophysical methods that are used in fractured rock investigations, GPR is arguably the most favorable tool to image individual fractures away from boreholes. 
Seismic methods, for instance, typically employ sources with a resulting larger wavelength and the acoustic impedance contrast is less pronounced than the electromagnetic impedance in most fractured rock environments \cite[e.g.,][]{Mair1981,Khalil_1993}. The resolution of electrical resistivity tomography is insufficient to identify individual fractures \cite[]{day2005applying,robinson2013evaluation,slater97}, but this is also the case for many other types of geophysical data when interpreted by smoothness-constrained inversion (such as seismic or GPR attenuation or travel time tomography; e.g., \cite{Day-Lewis_2003,ramirez1986investigation,daily1989evaluation}).

Surface-deployed GPR can be used to image shallow dipping fractures \cite[e.g.,][]{Grasmueck_1996}, whereas vertical borehole acquisitions can be used to image steeply dipping fractures \cite[e.g.,][]{Olsson_1992}. Borehole GPR reflection data can either be acquired in single-hole or cross-hole mode \cite[]{Dorn_2012}. 
Common borehole GPR systems create omni-directional radiation patterns (symmetry around the borehole axis), which precludes the determination of azimuths of plane reflectors when using data from one borehole alone. Directional antennas that were originally developed for  the characterization of prospective nuclear waste repositories, have only recently become more widely developed, but are still not commonly used \cite[]{slob2010surface}. 

The deployment of GPR monitoring during saline tracer tests is useful to identify those fractures that are transmissive and connected to the injection point. Tracer movements can thus be imaged in individual fractures away from the borehole locations \cite[e.g.,][]{Talley_2005,Dorn_3rd}. In the following, we refer to this data type as time-lapse GPR data.

The accessibility of fractured rock formations are typically restricted to the boreholes. This restriction has important consequences, in that the orientation of fractures can be well resolved at the borehole locations (using optical or sonic televiewer logs), but only within significant uncertainty away from the borehole (e.g., using GPR reflection images) \cite[]{Olsson_1992, Dorn_2012}. Furthermore, hydrological and geophysical data generally constrain mainly hydraulic or geometrical properties, respectively. The integration of different data types can help to overcome some of the limitations commonly encountered when characterizing fracture networks. As both hydrological and geophysical data and their interpretations contain considerable uncertainties, they should be carefully taken into account within a formal data integration procedure \cite[e.g.,][]{legoc}.

The use of geophysical data in building aquifer models has been investigated thoroughly in recent years \cite[e.g.,][]{hubbard1997ground, gallardo2007joint, linde2006improved, day2000identifying, chen2006development}. Many successful field demonstrations of deterministic integration methods exist that determine hydraulic and structural properties of unconsolidated materials \cite[e.g.,][]{looms2008monitoring,doetsch2010zonation}, but fewer examples consider fractured rock systems \cite[e.g.,][]{Day-Lewis_2003, niklas2004evidence}.
Deterministic inversion algorithms seek one model that fits the data, while deviating the least from a reference model or preferred model morphology. Their applicability may be limited for highly non-linear problems as the typically used gradient-based optimization techniques may get trapped in local minima and they provide limited insights about parameter uncertainty \cite[e.g.,][]{tarantola2005inverse}. Probabilistic inversion methods are suitable when it is expected that very different models (e.g., in terms of fracture geometries) are consistent with the available prior information and data, as they may explore the posterior probability density functions (pdfs) \cite[e.g.,][]{mosegaard1995monte}. Probabilistic sampling-based approaches for high-dimensional problems (e.g., hundreds of parameters) are computationally expensive. This means that it is often necessary to decrease parameter dimensionality and their ranges as much as possible beforehand.

Numerous global search methods exist, for example Markov chain Monte Carlo methods (MCMC), simulated annealing or genetic algorithms. 
\cite{day2000identifying} conditioned the 3-D geometry of fracture zones to hydraulic data using a simulated annealing approach. MCMC has the advantage that parameter uncertainty is formally assessed \cite[e.g.,][]{hastings1970monte,mosegaard1995monte,laloy2012high}.
\cite{chen2006development} used MCMC sampling to estimate probabilities of zones having high hydraulic conductivities. Markov chain Monte Carlo methods generate parameter samples that converge to a stationary distribution that coincides with the posterior pdf under certain conditions \cite[e.g.,][]{hassan2009using, chen2006development}. 
Such methods can be inefficient for discrete fracture networks (DFN) for which small changes in fracture geometries might lead to very different model responses. 
In this case it may be efficient to use the brute force rejection sampling method, which is the only exact and also the simplest global search method \cite[]{mosegaard1995monte}. 

To investigate the 3-D heterogeneity of fracture networks, the stochastic generation of multiple DFN models \cite[e.g.,][]{Darcel_2003} that are conditioned to available data is appealing.
The conditioning of DFN models to borehole observations is common, but its utility has been questioned given the difficulty in correlating fracture aperture at the borehole locations to hydraulic apertures \cite[]{renshaw1995relationship} and the limited coverage of borehole data with respect to the investigated rock volume \cite[]{neuman2005trends}. DFN approaches rely on statistical descriptions of fracture distributions, but reliable statistics are often very difficult to obtain. To partly circumvent these problems, we propose herein to condition connected 3-D DFN models to hydraulic, tracer and GPR reflection data that are sensitive to fracture geometry and hydrological state variables. Also, the flow responses of the realized DFNs are numerically simulated. For the conditioning of the DFNs we use a hierarchical rejection sampling method. We apply our scheme to data from the Ploemeur site in France, where relatively few well-connected fractures appear to control the flow \cite[]{Dorn_3rd}. More specifically, we derive 3-D models of connected DFNs by combining (1) logging data together with (2) a few fundamental geometrical properties of connected fracture networks, (3) images of fractures obtained from GPR data acquired under natural conditions and during tracer test, (4) topological constraints that describe the order in which the tracer arrives in different fractures as inferred from time-lapse GPR data, and (5) hydrological information (e.g., tracer arrival times, massflux curves) inferred from tracer tests. The set of conditioned fracture models can then be used to evaluate fracture network statistics.

\section{Methodology}
Within this study, we generate and condition connected DFN models under the assumption of an impermeable rock matrix. Fractures are described as homogenous thin circular discs distributed in 3-D space, where each fracture is parameterized with a midpoint ($x, y, z$), a dip ($\varphi$), an azimuth ($\theta$) and a radius ($R$). The same uniform transmissivity is assigned to every fracture. 
 The geometry of all individual fractures in a given model is defined by a vector {\bf m}. A series of observed hydrological and geophysical data ${\bf d}_{\rm obs}$ stems from the noise-contaminated response of a true and unknown system.
The goal of the probabilistic data integration or joint inversion procedure is to draw samples from a pdf $p({\bf m})$ describing prior constraints on the model space that also agree with ${\bf d}_{\rm obs}$ within the associated data and modeling errors.
 The agreement between calculated data ${\bf d}_{\rm cal}$ and observed data ${\bf d}_{\rm obs}$ is evaluated by a likelihood function $L({\bf m}|{\bf d}_{\rm obs})$. Bayes� theorem is used to calculate the posterior pdf $p({\bf m}|{\bf d}_{\rm obs})$ given the prior $p({\bf m})$ and the likelihood function $L({\bf m}|{\bf d}_{\rm obs})$, where $L({\bf m}|{\bf d}_{\rm obs})$ equals $p({\bf d}_{\rm obs}|{\bf m})$,
\begin{equation}\label{bay}
p({\bf m}|{\bf d}_{\rm obs}) = const \text{ } p({\bf m}) L({\bf m}|{\bf d}_{\rm obs}) 
\end{equation}
where $const$ is a normalization constant \cite[e.g.,][]{mosegaard1995monte}. Many different methods have been developed to derive $p({\bf m}|{\bf d}_{\rm obs})$. 
Rejection sampling is the only exact Monte Carlo sampling technique and every accepted model is a random and independent draw from $p({\bf m}|{\bf d}_{\rm obs})$. This simple method draws proposals directly from the prior distribution and their acceptance is decided by the value of the likelihood function alone, that is, there is no comparison with the likelihoods of previously sampled models, as in MCMC. 

A suitable metric is needed to compare simulated and observed data for a proposed model. Depending on the data type, we use the infinity norm $l_{\infty}$ or the L2-norm $l_2$ norm. The $l_{\infty}$ norm is applicable if the error bounds are strict \cite[]{marjoram2003markov}. In this case, a model is accepted if all the residuals are within the error bounds of the observed data. The likelihood function is thus either zero or one. It may have complex patterns for fracture networks whose connectivity may change significantly with only small perturbations in fracture geometry. 

Different simulations are used to evaluate the model responses of each data type and the associated computational costs are typically very different (e.g., flow simulations are computationally more expensive than evaluating hydraulic connections). To keep the computational costs as low as possible, we apply a hierarchical formulation of Bayes� theorem \cite[e.g.,][]{glaser2004stochastic}, in which the samples from the posterior pdf at one conditioning level are subsequently evaluated at the next level using other data types. Time-consuming simulations of proposed models are thus only performed on those that agree with the data types for which quick simulations are available. 

Rejection sampling can be very slow when dealing with large data sets with small errors as the acceptance rate may be  prohibitively low. This algorithm applies therefore primarily to instances when there are comparatively few data and when data or modeling errors are rather large. We argue that this is often the case in fractured rock investigations. Furthermore, comparatively narrow bounds on the  prior distribution are crucial to make the rejection sampling approach feasible. For example, it is most useful to know how many fractures are part of the connected DFN and their approximate positions. To make this approach feasible, it is therefore necessary to use site-specific data to derive a bounded $p({\bf m})$ (these data cannot be used further as conditioning data in the likelihood function). Outmost care is needed to assure that these bounded priors reflect realistic distributions, thereby, avoiding over-confident predictions or incompatibilities between the prior and the likelihood terms. In this work, the properties of individual fractures are largely constrained by the prior bounds and subsequent conditioning is mainly used to assure that the proposed models are in agreement with data that characterize the behavior of the whole fracture network (e.g., hydraulic cross-hole connections) that are difficult to include in $p({\bf m})$.

To demonstrate the methodology, we consider a set of classical hydrological data: optical televiewer logs, flowmeter and tracer test data. Furthermore, we use single-hole GPR data that provide geometrical constraints on individual fractures that are part of the connected fracture network \cite[]{Dorn_2011,Dorn_3rd}. Indeed, the method capitalizes through its model parameterization on a detailed description of the fracture geometry derived from GPR imagery and televiewer data. The different data types and their uncertainties are described in the following.

(1) Optical televiewer logs constrain the intersection depth, dip and azimuth of fractures that intersect the boreholes. The uncertainties of these estimates are mainly related to the interpreters identification of the fractures, the depth positioning error of the tool (cm-range) and the accuracy at which the borehole deviations are determined (about 0.5$^\circ$ for the inclination, and 1$^\circ$ for the azimuth). Furthermore, it is difficult to assess from borehole logs alone how well the local dip and azimuth at the borehole-fracture intersection represent the mean orientation of the entire fracture plane. In granite formations, \cite{Olsson_1992} and  \cite{Dorn_2012} find similar dips of fractures identified at the borehole locations and fractures imaged by GPR reflections.

(2) Hydraulic and tracer tests are generally used to identify and hydraulically characterize the transmissive fractures that intersect boreholes (and implicitly the connected fracture network). Cross-hole flowmeter testing permits inference of hydraulic connections. Besides the resolution of the flowmeter tool (e.g., 0.3 m/min for impeller flowmeters), additional uncertainties are related to the imposed hydraulic gradient and the influence of natural flow. Massflux curves acquired at the different outflow locations during tracer tests are useful to infer the individual contribution of each flow path. Concentration measurements depend on the accuracy of the logger and calibration data, but experimental conditions often pose further limitations.

\begin{figure}[h]
 \center
 \includegraphics[width=0.8\textwidth,angle=0]{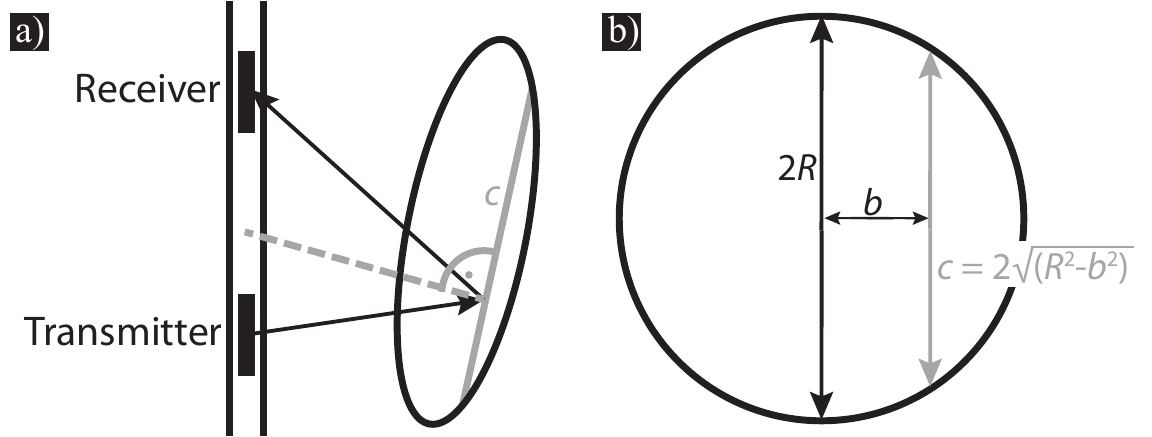}
 \caption[Schematic of circular fracture plane]{(a) Schematic of a simplified circular fracture plane with radius $R$ and the chord (in grey) on which normal vectors may exist that intersect the borehole (depending on the azimuth and the location of the fracture relative to the borehole) and thus allow the fracture chord to be imaged with GPR. (b) Plane view of fracture. The distance $b$ is undetermined.}
 \label{figchord}
\end{figure}

(3) Single-hole GPR reflection imagery provides very useful constraints on (i) the spatial extents of fractures that intersect a borehole (with intersection points determined from televiewer data) and (ii) geometrical attributes of fractures that do not intersect a borehole \cite[e.g.,][]{Olsson_1992, Dorn_2012}. Commonly used omni-directional GPR antennas provide 2-D projections (there is a circular symmetry along the borehole axis) of reflectors corresponding to fracture chords. These chords represent the parts of a fracture where normal vectors exist that cut the observation borehole location (Figure~\ref{figchord}). Due to the dipolar-type radiation patterns of GPR antennas, in which most of the energy is radiated normal to the borehole axis, GPR imagery is practically limited to fractures dipping between 30$^\circ$ and 90$^\circ$ for vertical observation boreholes. Fractures are typically observed in the range of few meters to some tens of meters away from the borehole, depending on the antenna frequency used and the electrical conductivity of the host rock. 
No information is obtained about the first 2 meters around the borehole, as this region is strongly dominated by the direct wave that masks early reflections from close-by fractures. 

The positioning accuracy of a fracture from its imaged fracture chord is affected by uncertainties of the GPR velocity used for migration, the picking error when interpreting reflections from migrated images ($\sim$ a quarter wavelength), the size of the first Fresnel zone (depends on the radial distance $r$ from the borehole and the signal frequency), and how far the midpoint of the imaged chord lies away from the midpoint of the fracture plane (distance $b$ in Figure~\ref{figchord}). 
For a fracture identified in a migrated single-hole GPR image, we define uniform prior distributions on the minimal and maximal extent in depth $z$ and radial distance $r$, which implies that the dip angle is indirectly defined. 
The positioning uncertainties of GPR-imaged fractures that intersect one borehole are different. 
For those fractures, we assign uniform prior distributions on the dip and strike angles, depth of borehole intersection, and its minimal and maximal extent in depth. For more details, see \cite{Dorn_2012, Dorn_3rd}.

\section{Field application}

\subsection{Field site} 
We apply our hierarchical rejection sampling scheme to data acquired at the well-studied Ploemeur aquifer test site in Brittany, France \cite[]{LeBorgne_2006, leray}, which is a long-term observatory for hydrogeological research. Our aim is to derive a relatively large set of connected DFN models that all describe observations in the vicinity of two ~6 m spaced boreholes B1 (83 m deep) and B2 (100 m deep). The geology consists of saturated granite overlain by 30-40 m of highly deformed mica schist. The granite formation, with low permeabilities of the intact rock matrix in the range of $3\times10^{-19}$ to $3\times10^{-20}$ m$^2$ \cite[]{Belghoul2007}, has the most permeable fractures \cite[]{LeBorgne_2007} and is the area of interest herein. 

\subsection{Available data}
\cite{LeBorgne_2007} find that the available formation at the experimental site is highly transmissive with overall hydraulic transmissivities on the order of $10^{-3}$ m$^2$/s over the length of each borehole. An ambient vertical pressure gradient resulting in an upward flow in the boreholes of about 1.5~L/min in each borehole is also expected to affect the flow in the fractures. From flowmeter and packer tests, \cite{LeBorgne_2007} identified the hydrologically most important fractures that are intersecting the boreholes and are part of the transmissive fracture network. There are only 4-5 such fractures intersecting each borehole over the thickness of the granite formation. In general, these fractures have dips in the range of 30-80$^\circ$ and azimuths in the range of 190-270$^\circ$. None of these identified fractures intersects both boreholes B1 and B2 \cite[]{LeBorgne_2007,Belghoul2007}. Possible hydraulic inter-borehole connections at the site inferred from a combination of single packer tests and cross-hole flowmeter tests are presented by  \cite{LeBorgne_2007}.

\begin{figure}[h]
 \center
 \includegraphics[width=.9\textwidth,angle=0]{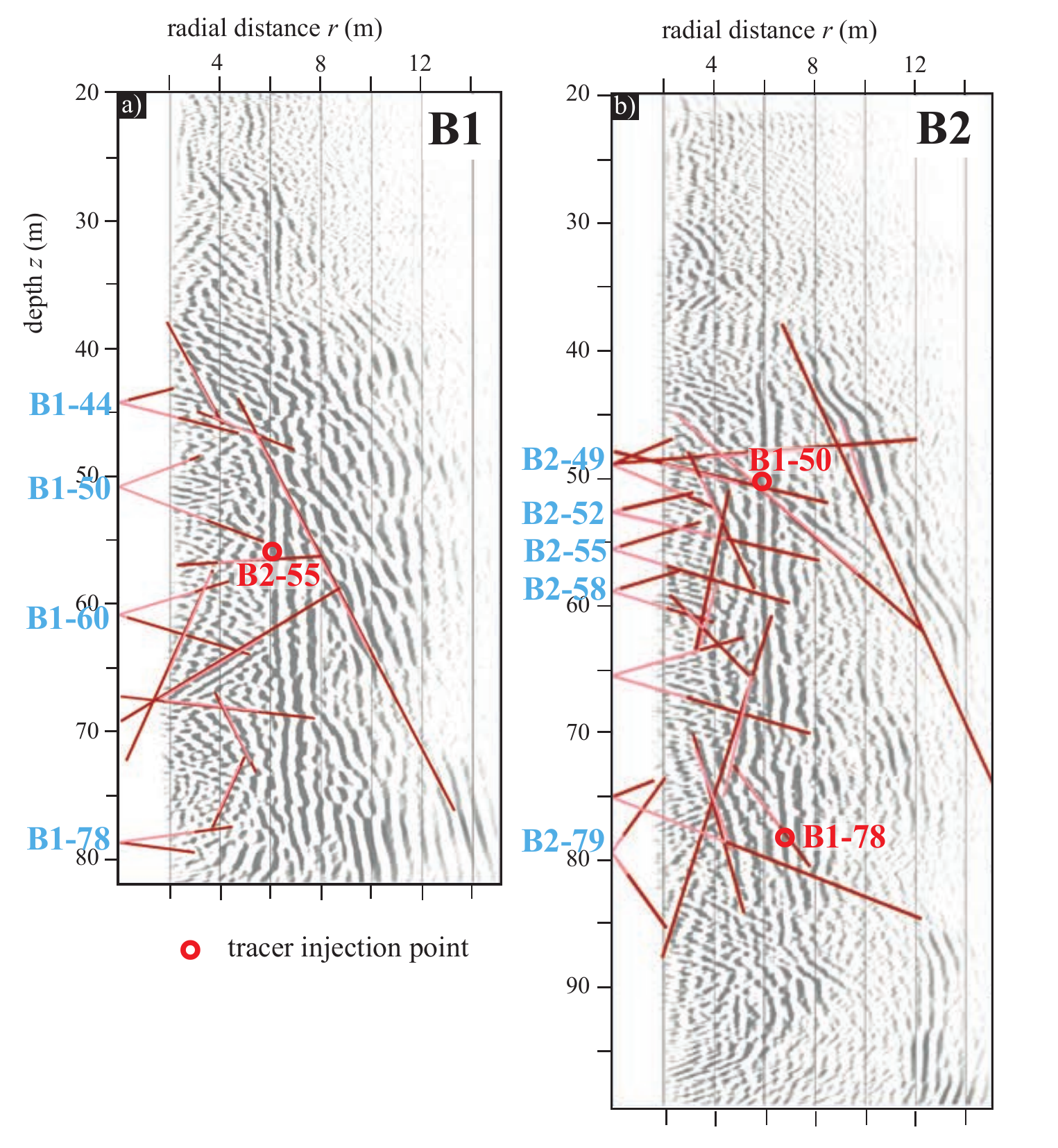}
 \caption[Interpretations of fractures based on GPR sections]{Extracts of the migrated multi-offset single-hole GPR sections of (a) B1 and (b) B2 from \cite{Dorn_2012} with superimposed interpretations of 27 fractures that are either inferred to be part of the connected fracture network using tracer tests and geophysical monitoring \cite[]{Dorn_2011,Dorn_3rd} or to be hydraulically important based on hydraulic tracer tests by \cite{LeBorgne_2007}. Light red colors depict the minimal extent of fracture chords, whereas dark red colors depict the maximal extent. Blue letters refer to hydraulically important fractures that intersect boreholes. The prominent vertical reflection is generated from the adjacent borehole (a) B2 and (b) B1 \cite[]{Dorn_2012}. Red circles refer to the location of the injection fractures used in the tracer tests by \cite{Dorn_3rd}. The axis aspect ratio $r:z$ is $2:1$.}
 \label{figpick}
\end{figure}

With the objective of imaging the local fracture network within the granite formation away from the boreholes, \cite{Dorn_2012} acquired 100 MHz and 250 MHz multi-offset single-hole GPR data. In the following, we refer to this data as static GPR data as they were carried out without imposing any perturbations  to the hydrological system by pumping or tracer injections. The majority of fractures that are identified as being transmissive by \cite{LeBorgne_2007} are imaged and geometrically characterized by \cite{Dorn_2012}. To identify among the imaged fractures those ones that are connected to the transmissive fracture network, \cite{Dorn_2011} and \cite{Dorn_3rd} performed single-hole GPR monitoring during saline tracer experiments. 
Three different transmissive fractures were chosen for the tracer injections (B1 at $z$ = 78.7 and 50.9 m and B2 at $z$ = 55.6 m) while pumping in the adjacent monitoring borehole. A total of 27 fractures are identified as being connected to at least one of the tracer injection points. We use the static GPR data from \cite{Dorn_2012} to constrain the geometry of these fractures (Figure~\ref{figpick}). They have a dip range of 30-80$^\circ$, with the lower limit being imposed by the detection limit of the GPR system. Yet, it is likely that the granite formation has very few subhorizontal fractures (dips below 30$^\circ$), because the borehole logs that are the most sensitive to subhorizontal fractures only indicate one transmissive fracture with a dip below 30$^\circ$ (in B1 at $z$ = 78.7 m dipping 15$^\circ$). 
During the course of the GPR-monitored tracer tests, the borehole fluid electrical conductivity $\sigma_w$ and hydraulic pressure $p$ were recorded in the monitoring borehole. Electrical conductivity profiles measured at different times were used to calculate concentration profiles and, together with previously acquired flowmeter data \cite[]{LeBorgne_2007}, to identify outflow locations and estimate massflux curves and mass recoveries at each such location (Table~\ref{tableone}).
The pressure conditions were rather unstable during these experiments, which imply that the pressure variations significantly influence the shape of the massflux curves \cite[see Figure 11 in][]{Dorn_3rd}. The total mass recoveries are also relatively low in all experiments ($\sim$15\%), which is likely due to the important natural flow gradient. The relative ratios of the total mass recoveries at the different outflow locations are used as conditioning data. Variations in the hydraulic head gradient are assumed to similarly influence the mass recoveries at the different outflow locations.

\subsection{Prior distribution}
The data used to define the prior distribution of connected DFNs are listed in the following. 

(1) Nine hydraulically important fractures are identified both in the borehole logs and in the GPR data, with three of them being used as tracer injection locations (Table~\ref{tableone}). Another 18 fractures are solely identified by GPR time-lapse data as they do not intersect the boreholes. Three out of these 18 fractures are imaged in GPR sections acquired in both boreholes B1 and B2, see \cite{Dorn_2011} for a discussion of how fractures are imaged from different boreholes. The total of 27 fractures are included in each proposed connected DFN.
Note that this number constitutes a minimum number of the actual fracture network, as we do not consider fractures outside of the detectable GPR-ranges.

(2) We impose that all 18 fractures that are identified solely from the time-lapse GPR data have to be connected to the corresponding fracture in which the tracer injection took place. They do not have to be part of the backbone of the hydraulic connections (i.e., the fractures through which the most significant fluid flow occurs) as the natural gradient or injection pressure may have pushed tracer into fractures that are not part of the backbone of the specific hydraulic connection. 
Nearly all fractures have to be connected, because some fractures play a role in several experiments (e.g., B2-55 plays a role in all three tracer tests; see Table~\ref{tableone}). The only fracture that is unconnected to the other fractures is B1-44. This fracture is identified by \cite{LeBorgne_2007}, but not in the time-lapse GPR and tracer test data by \cite{Dorn_3rd}. This fracture is included as its spatial extent restricts the feasible geometries of nearby fractures that must be unconnected to this fracture. 

(3) Optical televiewer logs constrain dips, azimuths and intersection depths of fractures that intersect boreholes. We account for an uncertainty of 5$^\circ$ and 3$^\circ$ in azimuths and dip angles. The directional uncertainty of the optical tool is much smaller (0.5$^\circ$). The larger values chosen are an attempt to account for the local fracture azimuths and dips in the borehole samples not necessarily being representative of these properties at the fracture scale.

(4) Fractures identified by time-lapse GPR data \cite[]{Dorn_2011,Dorn_3rd} are geometrically characterized by the static GPR data \cite[]{Dorn_2012}. We account for a uniform distribution within the following uncertainties:

\begin{itemize}

\item picking error of reflectors in the migrated GPR sections is considered to be a quarter of a wavelength (with a central frequency of 140 MHz and a GPR velocity of 0.12 m/ns, the wavelength is $\lambda$~=~0.86 m and the reading error is thus $\sim$0.2 m),

\item uncertainty of GPR velocity used in the migration ($\Delta v~=~\pm~0.002$~m/ns for the mean velocity of 0.12 m/ns based on tomographic inversion \cite[]{Dorn_2012}, which implies a relative error of $\Delta r/r = 2\times \Delta v/v = 4\%$ for the radial distances $r$ as the waves travel from the source to the reflector and then back to the receiver, 

\item	uncertainty of depth positioning of the final stacked GPR section is estimated to be $\Delta z = \pm 0.07$ m \cite[]{Dorn_2012}, 

\item	uncertainty due to the size of the Fresnel zone (depending on the radial distance $r$ of a feature, we add an uncertainty $\Delta z$ of half of the Fresnel zone that is given by $\nicefrac{1}{2} \text{ } sin(\varphi) \sqrt{\nicefrac{1}{2} \lambda r}$), where $\varphi$ is the dip of the reflector relative to the borehole. ($\Delta z \sim 1.5$~m for a reflector at $r = 5$ m, $\varphi = 90^\circ$ and $\lambda = 0.86$~m)

\item	uncertainty of the offset between the actual fracture diameter and the imaged chord $b$, that is chosen to vary between 0 and $0.75 \times R$ (see Figure~\ref{figchord}).
\end{itemize}

(5) Azimuths of fractures are only considered to be between 90-120$^\circ$ and 150-330$^\circ$, as all identified fractures in the televiewer data have azimuths in these ranges \cite[]{Belghoul2007}.

(6) Borehole deviation data define the relative distances between boreholes B1 and B2, and thus allows defining all fracture geometries in one common coordinate system.

(7) No fracture is allowed to intersect both boreholes. The 21 fractures identified in the time-lapse GPR data can only intersect an adjacent borehole if there is an open fracture observed at a similar depth of intersection ($\pm 0.15$ m), similar dip ($\pm 20^\circ$) and azimuth ($\pm 20^\circ$) as in the optical borehole logs.

\subsection{Conditioning data}\label{conditioning}
The hierarchical rejection sampling scheme is applied to three different conditioning levels that are described in the following.

(1) Conditioning level 1: Table~\ref{tableone} specifies eleven cross-hole hydraulic connections identified in previous tracer and cross-hole flowmeter experiments. These connections indicate pairs of transmissive fractures in neighboring boreholes that are connected or unconnected with each other. For each proposed model, we evaluate if the transmissive borehole fractures are connected with those of the neighboring borehole through the proposed DFN models and compare these connections with the experimental evidence in Table~\ref{tableone}. Only models for which all the connections are correctly predicted are retained ($l_{\infty}$ norm).

\begin{figure}[h]
 \center
 \includegraphics[width=1\textwidth,angle=0]{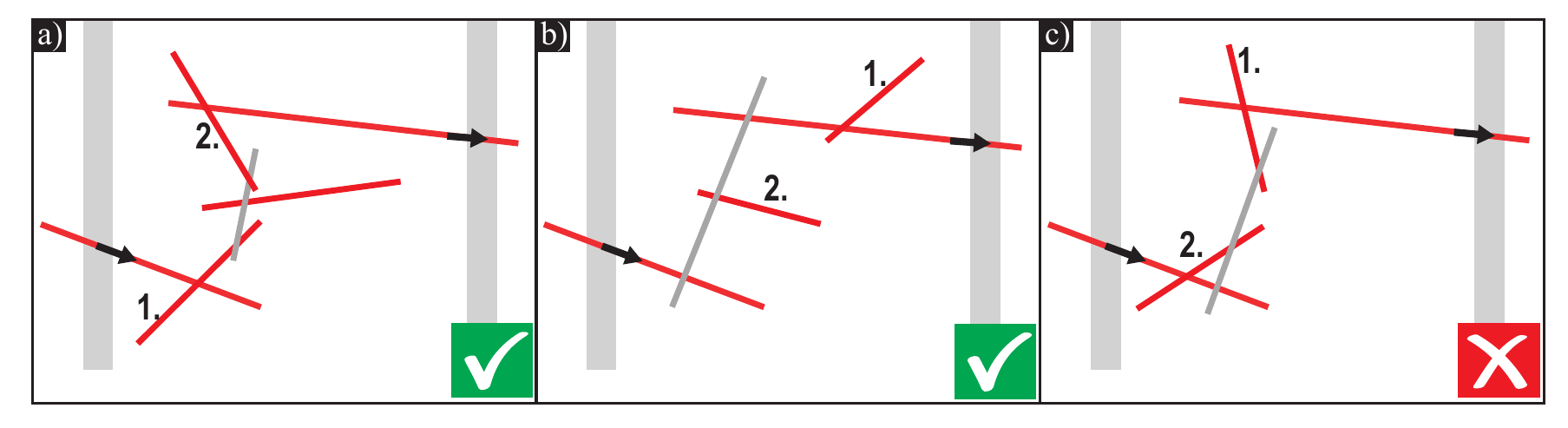}
 \caption[Schematic on acceptance and rejection: Conditioning level 2]{Simplified 2-D schematic describing (a-b) acceptance and (c) rejection of fracture models at conditioning level 2. Red colored fractures are identified in the time-lapse GPR data to be connected to the injection location. Grey colored fractures are identified in other tracer tests. Both types of fractures can be part of a hydraulic connection between injection and outflow points of the neighboring observation boreholes. A fracture model is only accepted if the connection order of fractures (indicated by the numbers in the figure) in the backbone agrees with those inferred from the GPR time-lapse images. The model (a) is accepted because the two fractures are part of the backbone and they appear in the right order, (b) is accepted because fractures for which a connection order is assigned are not part of the backbone and (c) is rejected because the two fractures are part of the backbone and they appear in the wrong order.}
 \label{figyesno}
\end{figure}

 (2) Conditioning level 2: For all tracer experiments and connections (Table~\ref{tableone}), we compare the sequential order of fractures connecting the backbone between injection and outflow points (e.g., fracture A is followed by fracture B) with the order inferred from tracer arrival times interpreted from time-lapse GPR images. We are conservative in only imposing 13 connection constraints that are clearly resolved in the time-lapse data. 
We use a graph representation of the fracture network to calculate the order of fracture connections. Fractures within the graph are represented as nodes and fracture connections as edges. The fracture connections can be determined by traversing the graph from a given node to another specified node using a depth-first search algorithm \cite[]{tarjan1972depth}. Figure~\ref{figyesno} depicts three scenarios of acceptance and rejection depending on the fracture network topology ($l_{\infty}$ norm).

 (3) Conditioning level 3: We simulate flow through the proposed DFN to test if the simulated relative flows at the tracer arrival locations are matching the measured relative mass at those locations (Table~\ref{tableone}). The flow model considers only imposed flow from injection to ouput fractures and does not include any background flow. This choice simplifies the numerical model. It implies that we are not able to simulate absolute outflow to the borehole, but we expect that the model provides good predictions of relative outflows. 
Flow is simulated using the Discrete Fracture Network model imbedded in the software H2OLAB \cite[]{poirriez2011etude,erhel2009flow}. Here, we consider homogenous and constant fracture transmissivities. The relative flow values will thus only be influenced by the topology and connectivity of the fractures. Estimations of flowpath transmissivities are discussed in section \ref{EST}. Within the fracture planes it is assumed that Darcy's law and mass conservation is satisfied. Only transversal flux is allowed on the fracture intersections (no longitudinal flux) and continuity of the hydraulic head and the transversal flux are imposed. For the injection and outflow locations at the fracture-borehole intersections, we impose Dirichlet boundary conditions and use Neumann zero flux conditions on fracture edges. The model is solved with a mixed hybrid finite element method that allows using locally refined meshes at fracture intersections. Within a fracture, the mesh is 2-D and at fracture intersections the two intersecting meshes are conform. The number of elements varies between $5\times 10^5$ and $1\times 10^6$, depending on the model realization.

The imposed head gradient corresponds to the mean gradient observed during the tracer experiments. The flow simulation does not account for the ambient vertical pressure gradient, on which we have only weak constraints. For this conditioning level, we use the $l_2$ norm to compare simulated flow ratios and observed mass recovery ratios. The estimated error of the observed mass recovery ratios is mainly influenced by the relative error of flow ($\sim15\%$ error) and the relative error of the measured concentration curves ($\sim20\%$ error). 
The accuracy of the conductivity logger (CTD diver~44077) is listed as 1\%, but the conductivity values are likely smeared out because the borehole fluid is mixed due to the up- and downward movement of the GPR antennas.

\section{Results}
\begin{figure}[h]
 \center
 \includegraphics[width=1\textwidth]{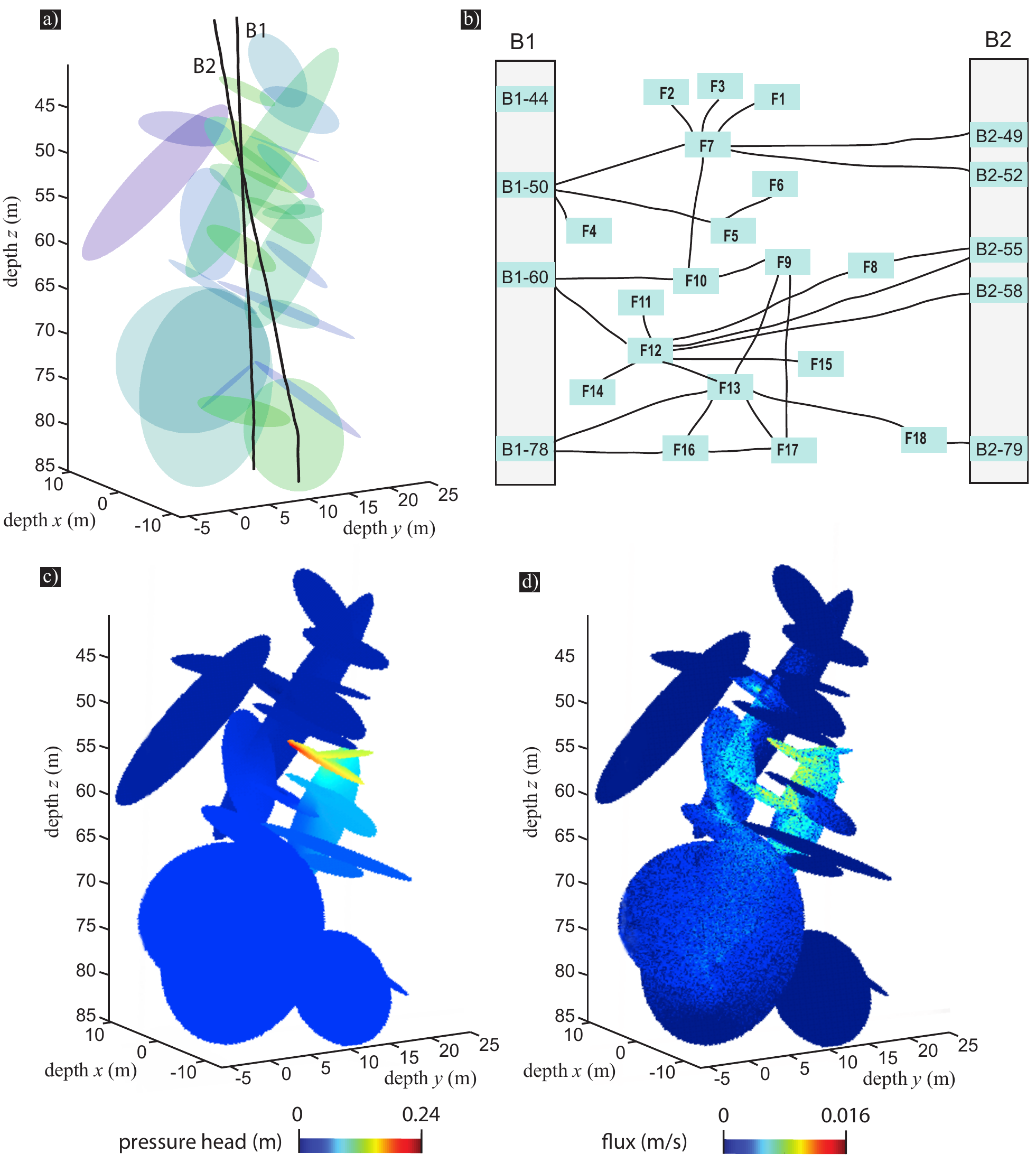}
 \caption[Example of fully conditioned DFN]{One realization of a fully conditioned fracture model (a) in 3-D Cartesian coordinates and (b) in graph representation comprising fractures that directly intersect the borehole (labeled Bx-x) and those that do not (labeled Fx). Modeled (c) hydraulic head and (d) flow distribution of the DFN realization for experiment III. The grainy look in (d) is exclusively related to the visualization. Note that there is a subtle difference in view angle in (a) compared to (b) and (c).}
 \label{figrealy}
\end{figure}

We use our hierarchical rejection sampling scheme to evaluate 2 million models drawn from $p({\bf m})$. Generating a single realization of a 3-D fracture model from $p({\bf m})$ requires, on average, 0.2 seconds of CPU time. A single fully conditioned DFN realization that is tested at all three conditioning levels requires, on average, 3 minutes of CPU time. Out of the tested models, 22\% honor the binary hydraulic connection information (conditioning level 1) out of which 40\% honor the GPR-constrained sequential order of the fractures (\mbox{conditioning level 2}). Out of the remaining models, only 0.1\% agree with the observed relative mass recoveries (conditioning level 3). A total of 230 of the proposed prior models honor all imposed data constraints. 
Figure~\ref{figrealy} shows different representations of one fully conditioned fracture model. Figure~\ref{figrealy}a is a spatial representation in Cartesian coordinates, whereas the graph representation in Figure~\ref{figrealy}b indicates the connections between the fractures. Figure~\ref{figrealy}c and d shows the modeled distribution of hydraulic head and flow of this model for experiment III (see Table~\ref{tableone}).

To evaluate the statistical properties of the networks at different conditioning levels, we use the following geometrical characteristics of the fracture networks: mean of fracture radii $\bar{R}$, the mean number of intersections per connected fracture $\bar{n}_{inter}$, and the mean length of fracture intersections. 
The mean of fracture radii is a key parameter to describe the connectivity of a fracture network \cite[]{bour1998connectivity}. 
The average number of intersections per fracture may be used to quantify the connectivity of a fracture network \cite[]{berkowitz1995analysis,robinson1999connectivity}. The mean intersection length quantifies if fracture intersections are well connected or not.  In the following, we discuss the different statistical measures in detail.

To compare the posterior and prior distributions, we use the relative information content ($RIC$) defined by \cite{tarantola2005inverse} as:
\begin{equation}
RIC = \sum_{i}{p_i({\bf m|d}) \log_{10} \frac{p_i({\bf m|d})}{p_i({\bf m})}}.
\end{equation}
At each conditioning level, the information content is expected to increase relative to the prior distribution. As an example, a gaussian distribution whose standard deviation is halved has a $RIC$ of $\sim$0.1. Note that the results of the RIC do not depend on the order in which the conditioning data are considered.

\subsection{Mean and variance of fracture radii}
\begin{figure}[h]
 \center
 \includegraphics[width=1\textwidth]{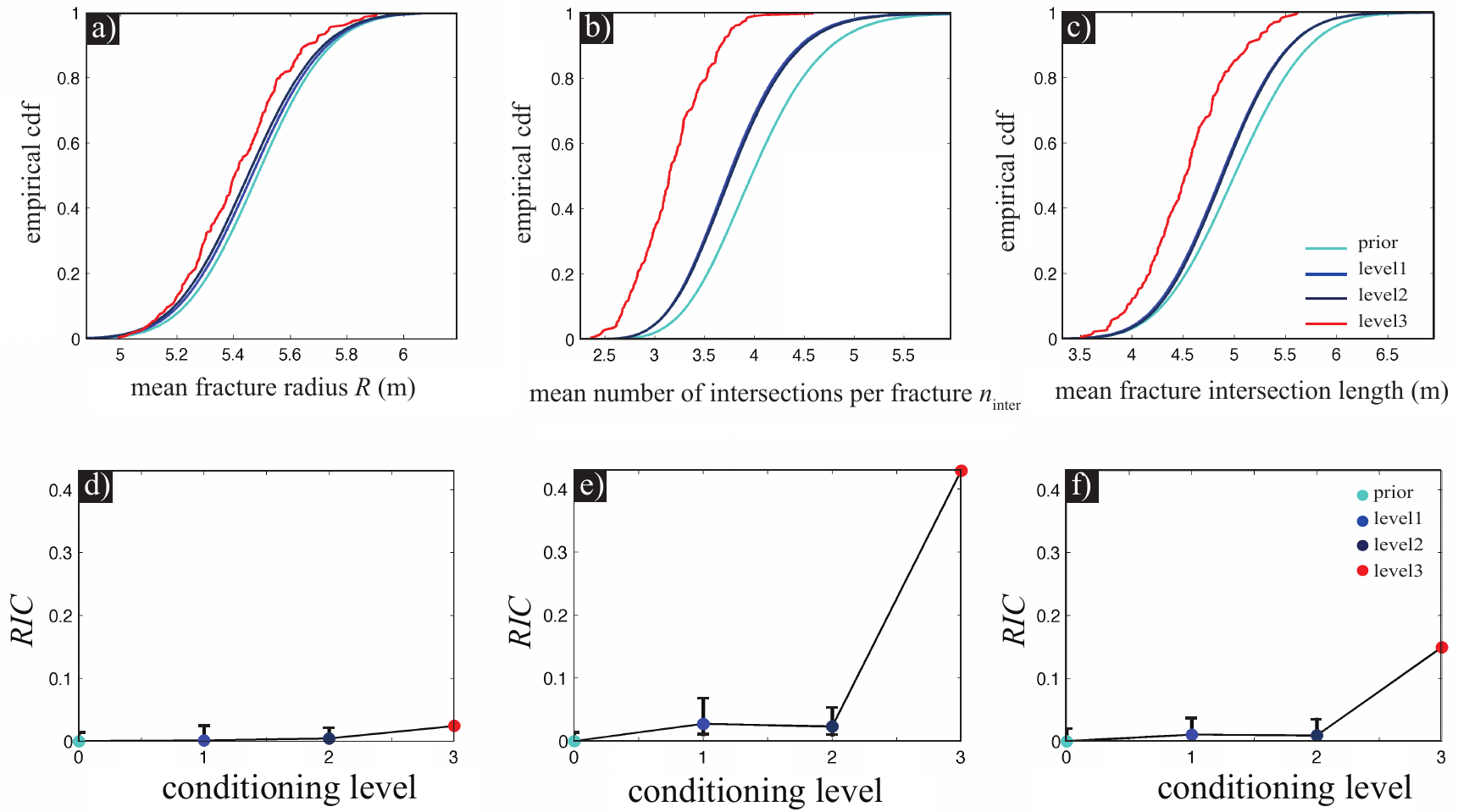}
 \caption[Summary statistics: Different length metrics]{Empirical cumulative distribution of (a) the mean fracture radius $\bar{R}$ and (d) its $RIC$, (b) the mean number of intersections per connected fracture $\bar{n}_{inter}$ and (e) its $RIC$, and (c) the mean fracture intersection length and (f) its $RIC$. The $RICs$ of conditioned distributions are given relative to $p({\bf m})$ at different conditioning levels. The  empirical cdfs in (a-c) are for the prior distribution based on $\sim 2 \times 10^6$ realizations, conditioning level~1 on $\sim 4\times 10^5$ realizations, conditioning level~2 on $\sim  2\times 10^5$ realizations and conditioning level~3 on $2.3\times 10^2$ realizations. The vertical bars in (d-f) indicate the variation of $RIC$ for randomly extracted subsets of the same sample size as condition level 3 (230 samples).
If the $RIC$ at condition level 3 lies outside of the ranges of $RIC$ at the lower conditioning levels, then the added information of level 3 is statistically significant.}
 \label{figcdf}
\end{figure}

Figure~\ref{figcdf}a shows the cumulative distribution function (cdf) of the mean fracture radius $\bar{R}$ for model realizations at different conditioning levels. The distribution of this parameter is well constrained in the prior (4.9~m $< \bar{R} <$ 6.1~m) by the GPR-constrained geometrical bounds on the fracture geometries. 
Mean fracture radius $\bar{R}$ (Figure~\ref{figcdf}a) of the prior set of models are on average higher than the mean fracture radius of the final conditioned DFNs. The difference between the 50th-percentile of prior and level~3 cdfs is about 39\% of the prior's standard deviation.
Figure~\ref{figcdf}d illustrates that the information content increases mainly from conditioning level 2 to 3.

\subsection{Number of intersections per connected fracture}
Similar to the mean fracture intersection radius $\bar{R}$, the mean number of intersections per fracture $\bar{n}_{inter}$ (Figure~\ref{figcdf}b) is related to the mean fracture area. We define this parameter by:

\begin{equation}
\bar{n}_{inter} = \frac{\bar{R^2 }}{N} \sum_{i}^N \frac{n_{inter,i}}{R_i^2},
\end{equation}
where $n_{inter,i}$ is the number of intersections of the $i$th fracture.
This measure shows clearly the impact of the conditioning with a difference between the 50th-percentile of prior and level~3 cdfs of 144\% of the prior's standard deviation. The largest difference between the conditioning levels can be seen between the models of conditioning level 2 and 3. The constraints of the flow response on the topology of the DFNs condition the models to smaller $\bar{n}_{inter}$, which in this case is related to a less dense fracture network and stronger exclusions between fractures.
Figure~\ref{figcdf}e indicates a clear increase in the information content when adding the conditioning data of level 1 and 3. The vertical bars for conditioning levels 1 and 2, indicate that the jump of $RIC$  from conditioning level~2 to 3 is statistically significant. 

\subsection{Mean fracture intersection length}
The mean fracture intersection length (Figure~\ref{figcdf}c) is the length on a fracture plane that is shared with another fracture. As for the mean fracture radius, the distribution of the mean fracture intersection length indicates lower values in the level~3 distribution than in the prior distribution, but the differences are more pronounced. Here, the difference between the 50th-percentile of prior and level~3 cdfs is 86\% of the prior's standard deviation. This larger difference is explained by the fact that the probability of a fracture intersection depends on the area of a fracture plane rather than the fracture radius. This implies that the variations in mean fracture intersection lengths are related to the square of the variations in mean fracture radius \cite[see also][]{de2000percolation}. 
Again, the cdf of the intersection length is well constrained by the GPR-constrained bounds on the prior geometries of fractures. The reason for the typically smaller mean fracture intersection length in the fully conditioned DFNs is the same as described above. The fully conditioned DFNs have fewer fracture interconnections and thus fewer fractures are realizing a hydraulic backbone connection.
Figure~\ref{figcdf}f again shows a statistically significant jump of $RIC$  from conditioning level~2 to 3.

\begin{figure}[h]
 \center
 \includegraphics[width=0.6\textwidth,angle=0]{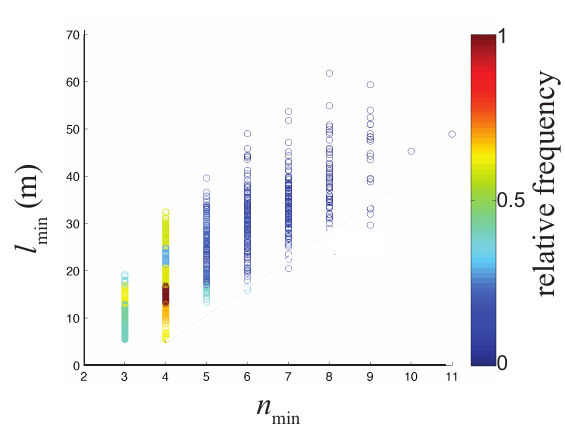}
 \caption[Number of fractures involved in hydraulic connection]{Relative frequency of the number of fractures $n_{\rm min}$ present in the connected fractures that establishes the minimum pathlength $l_{\rm min}$ for each of the cross-hole connections (Yes-connections in Table~\ref{tableone}).}
 \label{figscatter}
\end{figure}

Figure~\ref{figscatter} shows how many fractures $n_{\rm min}$ are needed for the fully conditioned models to establish the hydraulic connection with a minimal path length of $l_{\rm min}$, which is the shortest connection for a given hydraulic connection of a proposed DFN. The higher $n_{\rm min}$ for a given distance $l_{\rm min}$, the higher is the probability of having multiple flowpaths for a hydraulic connection. For example, if each of the hydraulic connections are realised by 5 fractures the minimum path length lies above 10 m. And there are at least 5 fractures involved if the minimum path lengths are longer than 33 m.

\subsection{Effective transmissivity estimation}
\label{EST}
In previous sections, we have focused on the relative flow contribution of flowpaths. The fully conditioned DFNs only differ in terms of fracture geometry and connectivity, while fracture transmissivity was set as a constant value for all fractures in the flow model. Absolute values of flowpath transmissivities can be estimated by rescaling the simulated outflow $Q_{\rm model}$ for an arbitrary transmissivity $T_{\rm model}$ to the flow that is estimated to be effectively occurring between injection and outflow fractures in the experiments. Recall that our model only simulates flow through fractures that connect the two wells and disregard other flow contributions that may come from other fractures. The occuring flow of a hydraulic connection, which we call $Q_{\rm tracer}$, may be estimated from field data by analyzing the dilution of tracer concentrations from the injection to the pumping well.
It is possible to estimate $Q_{\rm tracer}$ under the assumption that (1) dispersion is significantly smaller than dilution, and thus that the ratio of pumped tracer concentration $c_{\rm pump}$ versus injected tracer concentration $c_{\rm inj}$ is a reasonable approximation for dilution, (2) dispersion can be neglected, (3) pumping rates are stable throughout the experiments and (4) the tracer is conservative. Note that we do not account for the diffusion coefficient of the used tracer. The mass recoveries $m/m_0 =$ 15\%, 32\% and 32\% for experiments I, II and III, respectively, have also to be taken into account \cite[see][]{Dorn_3rd}. 
The resulting estimation of effective fracture transmissivity is
\begin{equation}
T_{\rm est} = \frac {Q_{\rm tracer}}{Q_{\rm model}} T_{\rm model}
\end{equation}
where $Q_{\rm tracer}$ is approximated by
\begin{equation}
Q_{\rm tracer} = Q_{\rm pumped} \frac {c_{\rm max}}{c_{\rm inj}} \frac {1}{m/m_0},
\end{equation}
where  $Q_{\rm pumped}$ is the pumping rate and $c_{\rm max}$ is the maximum tracer concentration of the measured tracer curve. 

\begin{figure}[h]
 \center
 \includegraphics[width=1\textwidth, angle=0]{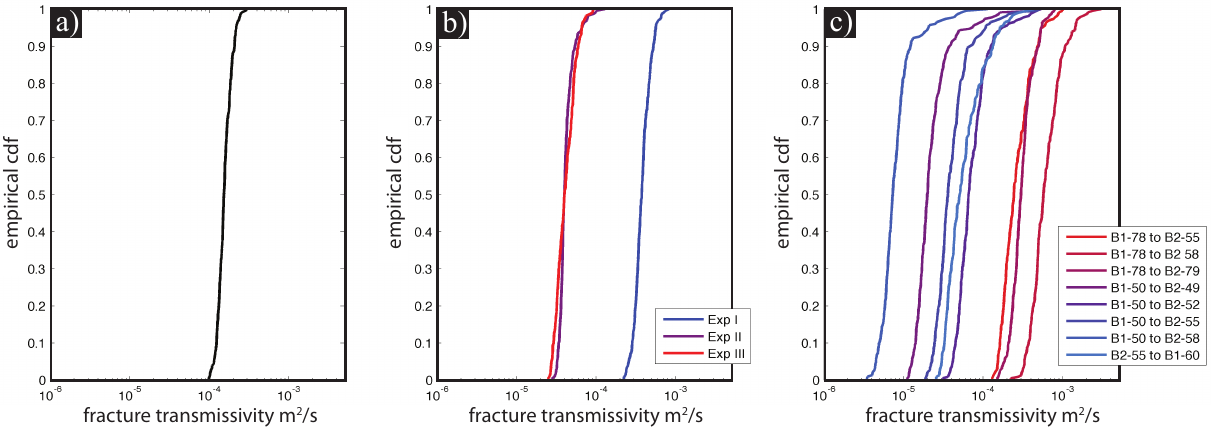}
 \caption[Transmissivity estimates]{Effective fracture transmissivity estimates $T_{\rm est}$. Estimates for (a) the entire network considering the total outflow of the DFN for all modeled experiments, (b) the entire network considering separately the total outflow of the DFN for the modeled experiments I, II and III. (c) Estimates of effective fracture transmissivity of the individual hydraulic connections by considering each individual outflow location separately.}
\label{figtrans}
\end{figure}

Note that the transmissivity estimates are performed on the set of fully conditioned DFNs, for which a uniform transmissivity was used during conditioning.
Figure~\ref{figtrans}a-c displays the estimated ranges of effective fracture transmissivities $T_{\rm est}$ on the set of fully conditioned models.
Effective fracture transmissivities for the entire network are ranging between $1-3\times 10^{-4}$ m$^2$/s (Figure~\ref{figtrans}a). This implies that differences in the geometry and the connectivity of the final DFNs are small in terms of their impact on the effective transmissivities. When considering every experiment separately, the effective transmissivities between experiments vary over one order of magnitude (Figure~\ref{figtrans}b). Estimates of the effective transmissivity of each hydraulic connection leads to variations in the range of $10^{-6}$ to $10^{-3}$ m$^2$/s (Figure~\ref{figtrans}c). This range matches well with estimates based on temperature tomography by \cite{marysoon} at the same site who estimated values between $2\times 10^{-6}$ and $8 \times 10^{-4}$ m$^2$/s.

\section{Discussion}

The proposed hierarchical rejection sampling inversion scheme yields samples from the posterior pdfs of connected DFNs at different levels of conditioning. The rather low acceptance rates at each level of conditioning indicate that new information is added. To quantify the added information, we analyze the empirical cdfs of average geometrical measures and their $RICs$. We find that the geometrical characteristics of the DFNs are already strongly constrained by the imposed geometrical constraints used to establish the prior $p({\bf m})$. Here, static and time-lapse GPR data are key for defining the prior models. Static GPR data provide geometrical constraints on the fractures within the GPR detection ranges \cite[]{Dorn_2012}. Depending on the media and the scale of observation, there are commonly numerous GPR-detected fractures. Time-lapse GPR data during specific tracer tests can be employed to determine which among those fractures that are connected and transmissive. This additional information makes it possible to dramatically decrease the number of model parameters and assume that the number of fractures are known. In our case study, the time-lapse data allowed reducing the numbers of hydraulically important fractures by a factor of five compared with the fractures identified by the static GPR. The corresponding reduction in the number of model parameters was necessary to make the rejection sampling feasible.

The final conditioned models are centered towards smaller mean and variance of fracture radii than the prior pdf (Figure~\ref{figcdf}). The uncertainty of the upper bounds on fracture radii caused by the underdetermined parameter $b$, is reduced by the conditioning data.
 The hydraulic connection data (conditioning level~1) has a small effect on the distributions. This is mainly because (1) the GPR-identified fractures of the proposed prior models are already connected (by definition) to the corresponding injection fractures and (2) there is only one fracture for which a hydraulic connection is not observed in the field (B1-44 in Table~\ref{tableone}). The acceptance rate of 22\% at conditioning level 1 is thus mainly related to the probability of fracture B1-44 being disconnected to the DFN. 
 The GPR-constrained sequential order of fractures (conditioning level~2) has some effect on the fracture topology. The prior distribution has a significant uncertainty in fracture radii and azimuths which leads to realizations with different fracture connections and thus to different network topologies. The hydraulic connection data and the sequential order of tracer arrival inferred from GPR data directly constrain the topology of the DFNs. The relative flow ratios indirectly constrain the topology of the DFNs as the interconnections of fractures and their geometries have an influence on the flow response. This conditioning (level~3) has the lowest acceptance rate and the largest effect on the conditioning of the topology of the DFNs.
 
The differences in the geometry of the fully conditioned DFNs only lead to small variations (a factor of three) in the effective transmissivities. An order of magnitude difference in transmissivity is found when estimating effective fracture transmissivities for the three tests individually (Figure~\ref{figtrans}b). For the individual hydraulic connections, the range in transmissivity estimates span 3 orders of magnitudes (Figure~\ref{figtrans}c). This indicates that the variability in geometry of the fully conditioned DFNs is less important than the heterogeneity in the transmissivities of individual fractures. In order to come up with models that can be used in a predictive sense, the variation in fracture transmissivities must be taken into account.

The assumptions made in this study regarding the fracture parameterization and its determination will influence the results, and some of them could be relaxed in future studies \cite[]{DornThesis}.
Most possible extensions would involve increased computational times as they require either more advanced forward modeling (e.g., transport modeling) or higher model dimensions (e.g., including variable number of fractures or fracture transmissivities). Within this study we do not consider transmissivity heterogeneity within the fractures. This heterogeneity can significantly repartition flow in the fractures as well as in the network. Furthermore, we use a fixed number of fractures, which simplifies the problem but also dismisses a larger variability in the conditioned DFNs. An extension to include additional unresolved fractures would be possible, but at the cost of higher computing times.

To constrain more parameters, data are needed that are sensitive to those parameters. The implementation of more complex models is thus a question of computational power, available data and data quality. All simulations presented herein are performed on one computer processor. Speed-ups based on rejection sampling scales linearly with the number of processors, which means that higher dimensions and transport modeling would be computationally feasible when using, say, 100~processors in parallel, which is common practice nowadays.

\section{Conclusions}

The properties of hydraulic cross-hole connections strongly depend on the 3-D geometry and topology of the local network of connected fractures. We present a novel approach to study such a network at the field-scale using hydrological and single-hole GPR reflection data. We use different data types (hydraulic, tracer, televiewer and GPR reflection data) to define prior bounds and to condition connected 3-D fracture models such that they agree with all available data. A hierarchical rejection sampling method is used to generate large sets of such realizations. The bounds of the prior distribution (the number of fractures, their positioning, orientation and spatial extent) are largely defined by the televiewer and the GPR data. Their use in defining the prior distribution makes the stochastic scheme computationally feasible as these data strongly reduce the set of possible prior models. In applying the methodology to field data  acquired in a fractured granite formation, we find models that can reproduce all of the conditioning data (e.g., hydraulic connections and their relative degree of connectivity, as well as the sequential order in which tracer moves through fractures in the backbone structure). We also perform flow simulations on the proposed DFNs to condition them to the observed relative degree of connectivity. From these fully conditioned models, we derive length scales and connectivity metrics distributions of the network. 
 We find that the most important conditioning data in the considered case are those related to the relative flow repartition within the DFN. The posterior realizations exhibit a significant variability in effective transmissivities between the individual hydraulic connections. Assumptions about the number and geometrical bounds of fractures might bias the derived statistics and they will partly be relaxed in the future. We conclude that the stochastic generation of DFNs that are conditioned to both hydrological and geophysical data is a powerful approach to study 3-D connected DFNs in the field. 
 
\section*{Acknowledgements}
We are very thankful to the four anonymous reviewers and Sarah Leray for her help with the numerical platform H2OLAB. This research was supported by the Swiss National Science Foundation under grants 200021-124571 and 200020-140390, the european project Climawat and the French National Observatory H+.


 \cleardoublepage
\section* {References} 
\bibliographystyle{elsarticle-harv}
\bibliography{dorn_thesis.bib}

\newpage

 \begin{table}[H]
 \caption[Specifications of hydraulic connections]{Information on specific cross-hole hydraulic connections observed during three different tracer tests between boreholes B1 and B2 by \cite{Dorn_3rd}. The name of a fracture (e.g., B2-55) indicates in which borehole (here, B2) and at which depth range (here, 55.0~m~$\le~z <$~56.0~m) a fracture intersects the borehole. The parameter $l_{\rm direct}$ indicates the distance between the injection and outflow locations, while hydraulic connection indicates if the two fractures are hydraulically connected. The relative mass recovery indicates the fraction of the total recovered mass arriving at a given outflow fracture.}
  \center
 \includegraphics[width=1\textwidth,angle=0]{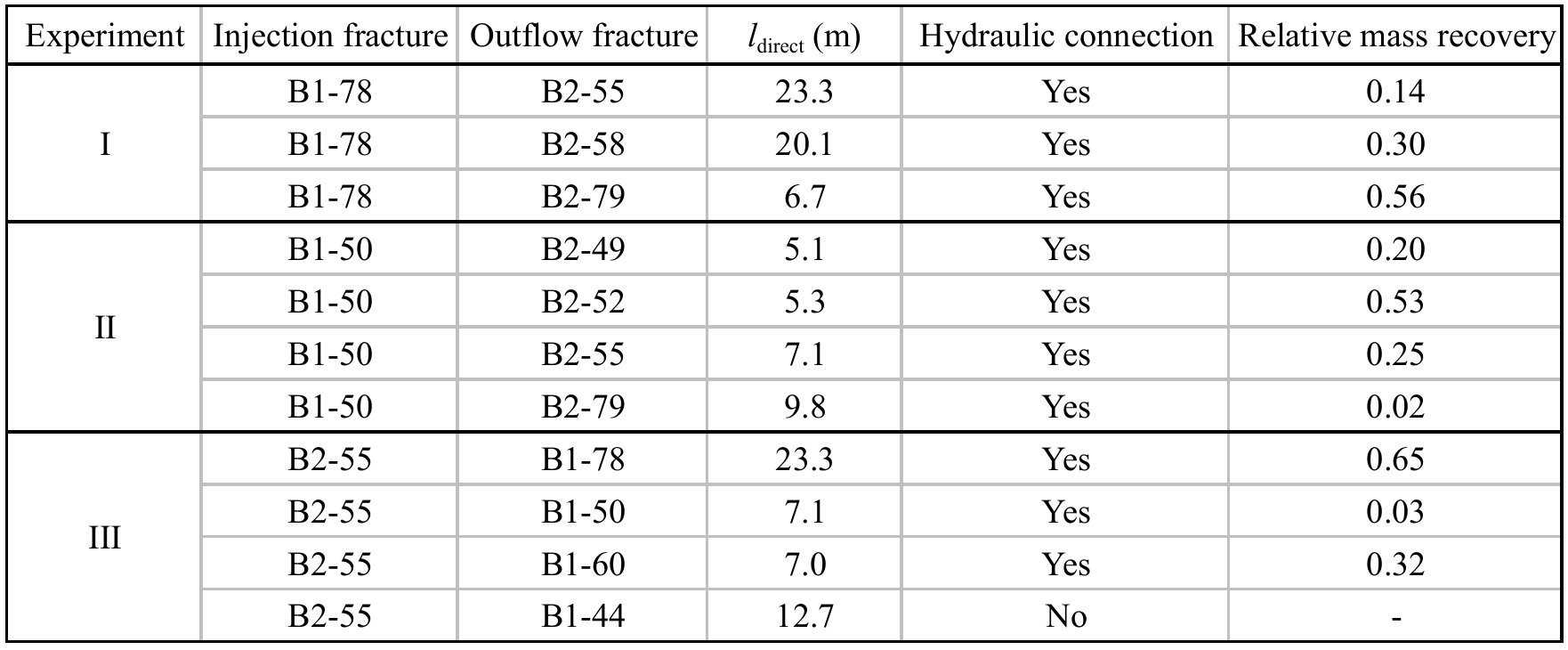}

 \label{tableone}
\end{table}

\end{document}